\documentclass[fleqn,12pt,twoside]{article}
\usepackage[headings]{espcrc1}

\readRCS
$Id: espcrc1.tex,v 1.2 2004/02/24 11:22:11 spepping Exp $
\usepackage{graphicx}
\usepackage[figuresright]{rotating}

\newcommand{\AmS}{{\protect\the\textfont2
  A\kern-.1667em\lower.5ex\hbox{M}\kern-.125emS}}

\title{Gross Properties and Isotopic Phenomena in Spectator Fragmentation}

\author{C.~Sfienti\address[GSI]{Gesellschaft f{\"u}r Schwerionenforschung, 
D-64291 Darmstadt, Germany},
M.~De~Napoli\address[INFNCT]{Dipartimento di Fisica dell'Universit\'{a} and INFN, I-95126 Catania, Italy}, 
P.~Adrich\addressmark[GSI], 
T.~Aumann\addressmark[GSI], 
C.O.~Bacri\address[FRA1]{Institut de Physique Nucl{\'e}aire, IN2P3-CNRS 
et Universit{\'e}, F-91406 Orsay, France}, 
T.~Barczyk\address[POL2]{M.~Smoluchowski Institute of Physics, Jagiellonian
Univ., Pl-30059 Krak{\'o}w, Poland}, 
R.~Bassini\address[INFNMI]{Istituto di Scienze Fisiche, Universit\`{a} 
degli Studi and INFN, I-20133 Milano, Italy}, 
S.~Bianchin\addressmark[GSI],
C.~Boiano\addressmark[INFNMI],
A.S.~Botvina\addressmark[GSI]\address[RUS]{Inst.~Nucl.~Res., 
Russian Accademy of Science, Ru-117312 Moscow, Russia},
A.~Boudard\address[FRA2]{DAPNIA/SPhN, CEA/Saclay, F-91191 Gif-sur-Yvette, France}, 
J.~Brzychczyk\addressmark[POL2], 
A.~Chbihi\address[GAN]{GANIL, CEA et IN2P3-CNRS, F-14076 Caen, France},
J.~Cibor\address[POL3]{H. Niewodnicza{\'n}ski Institute 
of Nuclear Physics, Pl-31342 Krak{\'o}w,Poland}, 
B.~Czech\addressmark[POL3],
J.-E.~Ducret\addressmark[FRA2], 
H.~Emling\addressmark[GSI], 
J.~Frankland\addressmark[GAN], 
M.~Hellstr{\"o}m\addressmark[GSI], 
D.~Henzlova\addressmark[GSI], 
K.~Kezzar\addressmark[GSI],
G.~Imm{\'e}\addressmark[INFNCT], 
I.~Iori\addressmark[INFNMI], 
H.~Johansson\addressmark[GSI], 
A.~Lafriakh\addressmark[FRA1], 
A.~Le~F{\`e}vre\addressmark[GSI], 
E.~Le~Gentil\addressmark[FRA2], 
Y.~Leifels\addressmark[GSI], 
W.G.~Lynch\address[MSU]{Department of Physics and
Astronomy and NSCL, MSU, East Lansing, MI 48824, USA}, 
J.~L{\"u}hning\addressmark[GSI], 
J.~{{\L}}ukasik\addressmark[GSI]\addressmark[POL3], 
U.~Lynen\addressmark[GSI], 
Z.~Majka\addressmark[POL2], 
M.~Mocko\addressmark[MSU], 
W.F.J.~M{\"u}ller\addressmark[GSI], 
A.~Mykulyak\address[POL1]{A.~So{\l}tan Institute for Nuclear Studies,
Pl-00681 Warsaw, Poland}, 
H.~Orth\addressmark[GSI], 
A.N.~Otte\addressmark[GSI], 
R.~Palit\addressmark[GSI], 
P.~Pawlowski\addressmark[POL3],
A.~Pullia\addressmark[INFNMI], 
G.~Raciti\addressmark[INFNCT], 
E.~Rapisarda\addressmark[INFNCT], 
H.~Sann\addressmark[GSI]\thanks{deceased}, 
C.~Schwarz\addressmark[GSI], 
H.~Simon\addressmark[GSI],
K.~S{\"u}mmerer\addressmark[GSI], 
W.~Trautmann\addressmark[GSI],
C.~Volant\addressmark[FRA2], 
M.~Wallace\addressmark[MSU], 
H.~Weick\addressmark[GSI],
J.~Wiechula\addressmark[GSI], 
A.~Wieloch\addressmark[POL2] and 
B.~Zwieglinski\addressmark[POL1] 
}

\runtitle{Gross Properties and Isotopic Phenomena in Spectator Fragmentation}
\runauthor{C.~Sfienti}

\begin{document}

% typeset front matter
\maketitle

\begin{abstract}
A systematic study of isotopic effects in the break-up of projectile 
spectators at relativistic energies has been performed with the ALADiN 
spectrometer at the GSI laboratory. \\
Searching for signals of criticality in the fragment production 
we have applied the model-independent universal fluctuations 
theory already proposed to track criticality signals in multifragmentation to our data. 
The fluctuation of the largest fragment charge and of the 
 asymmetry of the two and three largest fragments and their
  bimodal distribution have also been analysed.

\end{abstract}

\section{INTRODUCTION}
One of the most fascinating phenomena in physics is that of a phase transition.
Initially observed in macroscopic systems and in electromagnetic interactions, phase transitions have 
been seen manifesting also in strongly interacting microscopic systems and nowadays 
two specific areas are receiving a great deal of attention. One involves 
the loss of stability of excited nuclear systems which, under certain 
conditions of temperature and density, may lead to the total disassembly 
of the nucleus into particles and fragments. The second, at much higher 
energies, concerns the transition from hadrons to quarks and gluons, and the possibility 
of observing new phenomena in quark matter. \\
Nucleus-nucleus collisions at intermediate and relativistic energies have 
been shown~\cite{poc97} to be an ideal tool to produce pieces of finite nuclear 
matter at extremely different thermodynamical conditions. 
In order to analyze thermodynamical properties of microscopic systems we 
need to produce and select samples which can be associated with microcanonical ensembles 
at high excitation energies. \\
In a series of experiments~\cite{shutt}, 
multifragment decays of projectile spectators have been studied with the 
ALADiN forward-spectrometer at the SIS accelerator (GSI-Darmstadt-Germany). 
In these collisions, energy depositions are reached, which cover the range 
from particle evaporation to multifragment emission and further to the total 
disassembly of the system. 
The most prominent feature of the multifragment decay is the universality 
of the fragment multiplicities and the fragment charge correlations. 
The loss of memory of the entrance channel is an indication that 
equilibrium is attained prior to the fragmentation stage of the reaction. 
Clearly, new experiments are mandatory for having a better knowledge of 
the thermodynamics of a finite nucleus and its decay. \\
Recently a systematic investigation on projectile-spectator fragmentation 
has been undertaken at the ALADiN spectrometer at the GSI~\cite{sfie}: four 
different projectiles, $^{197}$Au, $^{124}$La, $^{124}$Sn and 
$^{107}$Sn, all with an 
incident energy of 600 AMeV on $^{nat}$Sn and $^{197}$Au targets, have been studied.

\section{CHARGE CORRELATIONS IN MULTIFRAGMENT DECAY}
\begin{figure}[htb]
\centering
\vspace{-1mm}
\begin{minipage}[c]{.49\textwidth}
   \centerline{\includegraphics[scale=0.5]{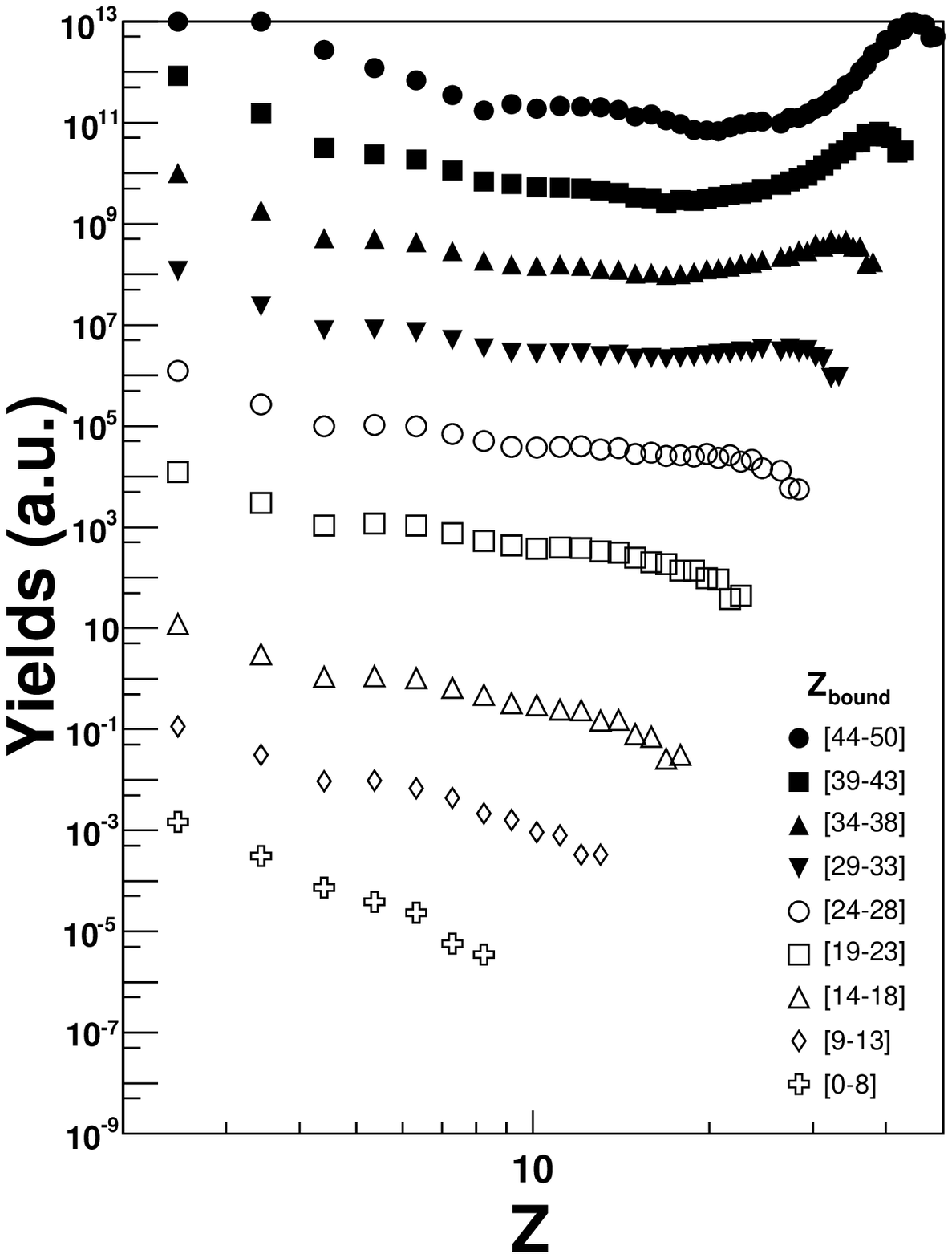}}
\end{minipage}
\begin{minipage}[c]{.49\textwidth}
   \centerline{\includegraphics[scale=0.4]{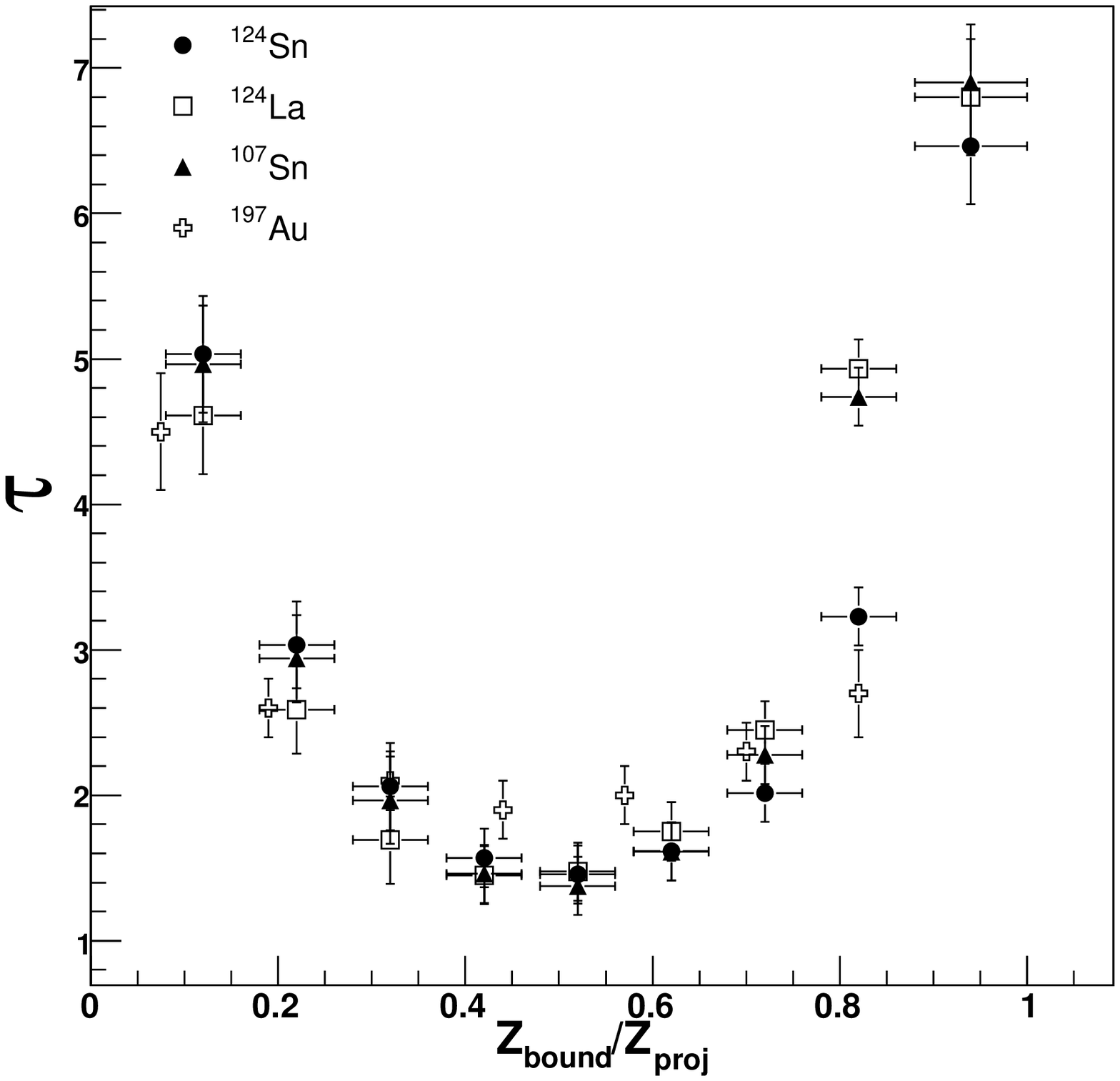}}
\end{minipage}
\vspace{-8mm}
\caption{Left Panel: Charge distributions obtained in the $^{124}$Sn fragmentation for different
 bin of $Z_{\rm bound}$.
Right Panel: The extracted $\tau$ parameters as a function of $Z_{\rm bound}$ for 
$^{124}$La, $^{124}$Sn and $^{107}$Sn at 600 AMeV, compared with the values of $^{197}$Au at
the same energy~\cite{kreutz}. 
}
\label{fig:charge}
\end{figure}

\begin{figure}[h]
\vspace{-5mm}
\centering
\includegraphics*[scale=0.45]{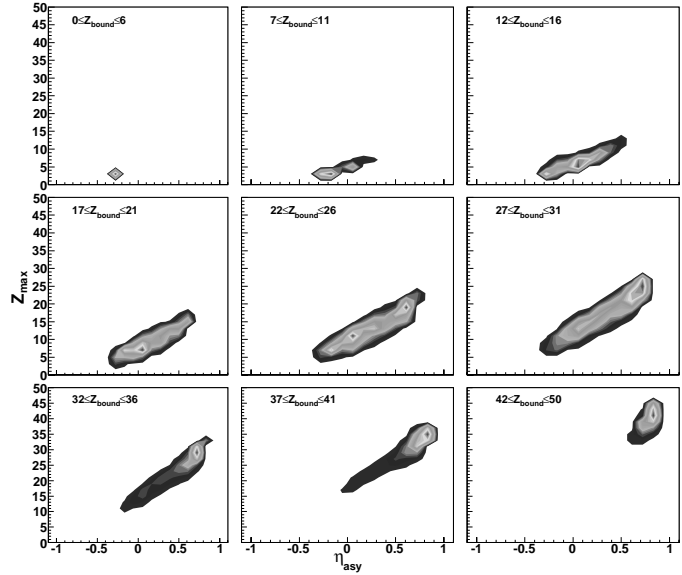}
\vspace{-8mm}
\caption{Event distributions in the $Z_{\rm max}$-$\eta_{\rm asy}$ plane for 
the ${\rm ^{124}Sn}$ projectile. Each panel refers to a different cut in $Z_{{\rm bound}}$.}
\label{bimod}
\end{figure}

The fragments emerging from the decay of the projectile spectators are well localized in rapidity. 
The distributions are concentrated around a rapidity value very close to the beam rapidity and become 
increasingly narrower with increasing mass of the fragment. The sorting out of the collision according 
to its centrality is done using the observable $Z_{\rm bound}$ defined as the sum of the charges 
of the projectile fragments ($Z>1$). To first approximation $Z_{\rm bound}$ is a measure of the charge 
of the remaining projectile-spectator except for the number of evaporated Z=1. It follows that large 
values of $Z_{\rm bound}$ correspond to big projectile-spectators, 
i.e. peripheral collisions, whereas small values of $Z_{\rm bound}$ correspond to small 
projectile-spectators, i.e. more central collisions. The excitation of the projectile-spectator 
depends on the impact parameter, since it occurs via knocking-out nucleons: 
$Z_{\rm bound}$ is therefore not only a measure of the impact parameter, 
 but also of the excitation energy. \\
In the left panel of fig.~\ref{fig:charge} the charge distributions measured in the fragmentation of 
$^{124}$Sn for 9 different 
bins of $Z_{\rm bound}$ are shown. The shapes of the charge distributions as a function of 
$Z_{\rm bound}$ are similar for all 
the studied systems. 
We have fitted these charge distributions with a power-law parameterization 
$\sigma (Z) \propto Z^{-\tau}$. The fitting range was $3 \leq Z \leq 15$.
The right panel of fig.~\ref{fig:charge} presents the evolution of the $\tau$ 
parameter as a function of $Z_{\rm bound}/Z_{\rm proj}$.
The $\tau$ parameter for all the three reactions lies on an almost universal curve and shows a 
minimum near $Z_{\rm bound}/Z_{\rm proj}$=0.5. 
Specific isotopic effects, even though small, can be nevertheless observed:
in particular, the inversion in the hierarchy of the three studied systems by going from small to high
values of $Z_{\rm bound}$ (the values for $^{124}$Sn are larger than those for the other two 
systems in central collisions and become slowly smaller than them by going towards larger impact 
parameters) has been shown to be related to an isospin dependence of the surface energy as a 
function of the excitation energy of the system~\cite{ogul}. \\
Since long it has been observed that some characteristics of the exited nuclear-system, 
such as the largest emitted 
charge ($Z_{\rm max}$) and the asymmetry between the three largest-charges
$\eta_{asy} = \frac{\left(Z_{\rm
max}-(Z_{2}+Z_{3})\right)}{\left(Z_{\rm max}+(Z_{2}+Z_{3})\right)}$
 vary by going from the evaporation to the vaporization regions through the multifragmentation 
regime. If the 
lowly-excited nuclear system evaporates light fragments, large $Z_{\rm max}$ and $\eta_{\rm asy}$ are expected, 
whereas in case of 
vaporization, both $Z_{\rm max}$ and $\eta_{\rm asy}$ are expected to be small. 
Fig.~\ref{bimod} shows the event distributions in the $Z_{\rm max}$ and $\eta_{\rm asy}$ plane for different 
cuts of $Z_{\rm bound}$ in the case of the $^{124}$Sn projectile.
For large values of $Z_{\rm bound}$, i.e. small excitation energies, the system is in the evaporation 
regime and therefore 
almost all the events
present high values of $Z_{\rm max}$ and an $\eta_{\rm asy}$ value near 1 (lower panels in fig.~\ref{bimod}). 
The distribution of $Z_{\rm max}$ as well as 
for $\eta_{\rm asy}$, is
characterized by the presence of a single peak. By going 
towards smaller values
of $Z_{\rm bound}$, the event distributions start to populate the region at intermediate 
$Z_{\rm max}$ and $\eta_{\rm asy}$ until, in the 
multifragmentation region,
events with high $Z_{\rm max}$ and $\eta_{\rm asy}$ and small $Z_{\rm max}$ and $\eta_{\rm asy}$ are simultaneously 
present (middle panels of fig.~\ref{bimod}). 
This is the region where the
two variables exhibit the largest fluctuations and correspondingly a bimodal distribution 
(central panel in fig.~\ref{bimod}). 
When, finally,
$Z_{\rm bound}$ is small enough, all the events are located in the vaporization zone 
(upper panels in fig.~\ref{bimod}) characterized 
by simultaneously
small values of $Z_{\rm max}$ and $\eta_{\rm asy}$. 
Similar behaviors have been found by analyzing other systems~\cite{iwm}. In particular this
analysis has also been extended to projectile-spectator systems studied by the
Aladin group in previous experiments~\cite{shutt} in order to explore a wider mass-range: for 
each system the occurrence of a bimodal distribution for a specific 
impact parameter has been observed. \\
Whether such results could be interpreted as a genuine bimodal signal characteristic of a 
first-order phase transition 
is still a matter of investigation: indeed the observation of a bimodal distribution in the data 
could be the result of a mixing of residue and multifragment emission for some impact parameter 
bins.

\section{$\Delta$-SCALING IN SPECTATOR FRAGMENTATION}
Signals of the predicted phase coexistence at 
low density and temperature ($\rho < \rho_{o}$ and $T < T_{c}$), often
associated with multifragmentation, may be revealed by abnormal fluctuations of experimental 
observables~\cite{fil1}. 
The theory of the
universal character of order parameter fluctuations in finite systems~\cite{bot} addresses 
this question in a 
model-independent framework, for systems with a second-order phase transition. 
Experimental observables that may be related 
to an order-parameter 
can be identified, in
particular through their $\Delta$-scaling behavior, and recently a model-independent tracking 
of criticality signals 
in nuclear
multifragmentation data at intermediate energies has been proposed~\cite{frank}. \\
\begin{figure}[h]
\vspace{-15mm}
\centering
\includegraphics*[scale=0.35,angle=-90.]{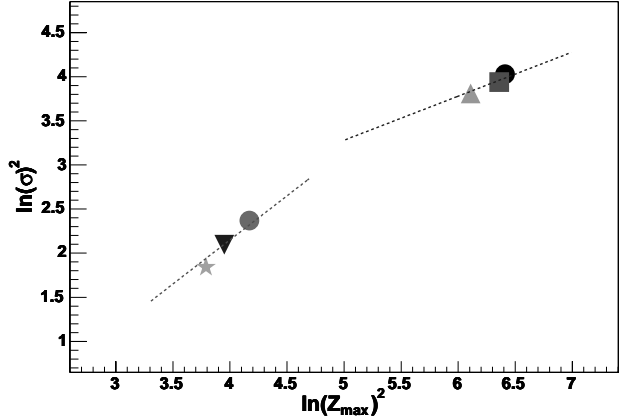}
\vspace{-12mm}
\caption{Correlation between the mean value and the fluctuation of
$Z_{\rm max}$: the various symbols refer to different cuts in the hodoscope multiplicity.}
\label{dsca1}
\end{figure}
The $\Delta$-scaling is observed when two or more 
probability distributions $P_{N}[m]$ of the observable {\em m} for 
a system size {\em N} collapse onto a single scaling curve $\Phi(z_{(\Delta)})$ 
independent of system size when plotted in terms of the scaling 
variables:
\begin{equation}
\left < m \right >^{\Delta} P_{N}\left [m\right]= \Phi(z_{(\Delta)}) = 
\Phi \left( \frac{m-\left < m \right >^{\Delta}} {\left < m \right >^{\Delta}}\right)
  \label{eqn:dscaling}
\end{equation}
where $\left<m\right>$ is the mean value of the distribution $P_{N}\left[m\right]$ and 1/2$\leq \Delta \leq$1. 
$\left<m\right>$ plays the role of a scale parameter and can replace {\em N} as a measure of the size of the
system. 
The scaling law with $\Delta$=1/2 is associated with low 
temperature ({\it ordered} systems), or with observables which are 
not related to an order parameter. Scaling with $\Delta$=1 is seen at 
high temperature ({\it disordered} system) and also for systems at critical conditions. 
A necessary condition for {\em m} to be an order parameter is that it must exhibit a corresponding change 
of $\Delta$-scaling regime when some suitable control parameter (e.g. available 
energy, temperature, etc.) is varied.\\
Using the data from this experiment, it has been investigated whether the $Z_{\rm max}$ 
distributions follow the $\Delta$-scaling law. 
The multiplicity of charged particles 
measured with the Catania Hodoscope~\cite{marzio} was used as impact parameter selector in oder to
avoid a possible autocorrelation of $Z_{\rm bound}$ with $Z_{\rm max}$. 
\begin{figure}[h]
\vspace{-15mm}
\begin{minipage}[t]{0.5\linewidth}
\centering
\includegraphics*[scale=0.33,angle=-90.]{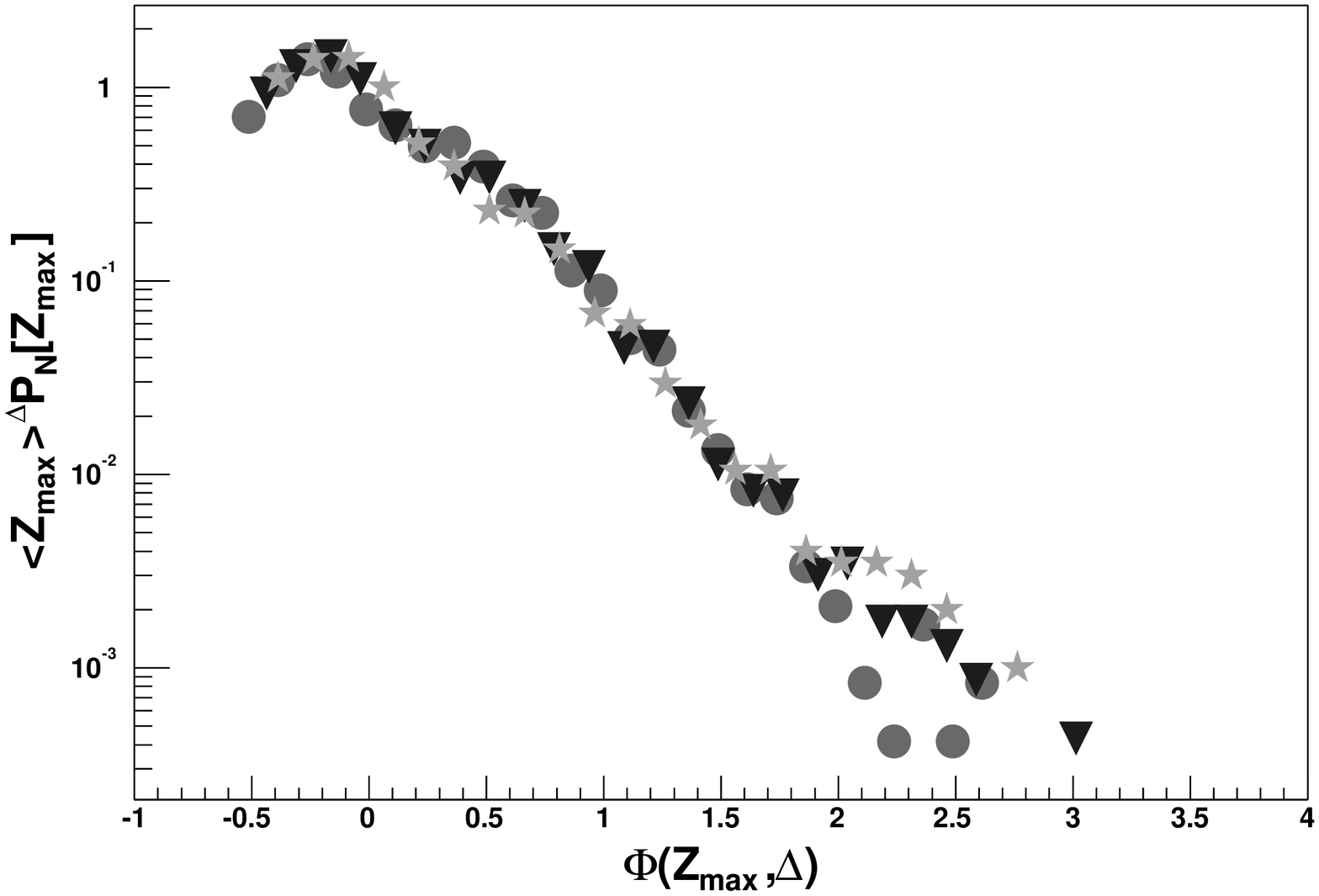}
\end{minipage}
\hspace{\fill}
\begin{minipage}[t]{0.5\linewidth}
\centering
\includegraphics*[scale=0.33,angle=-90.]{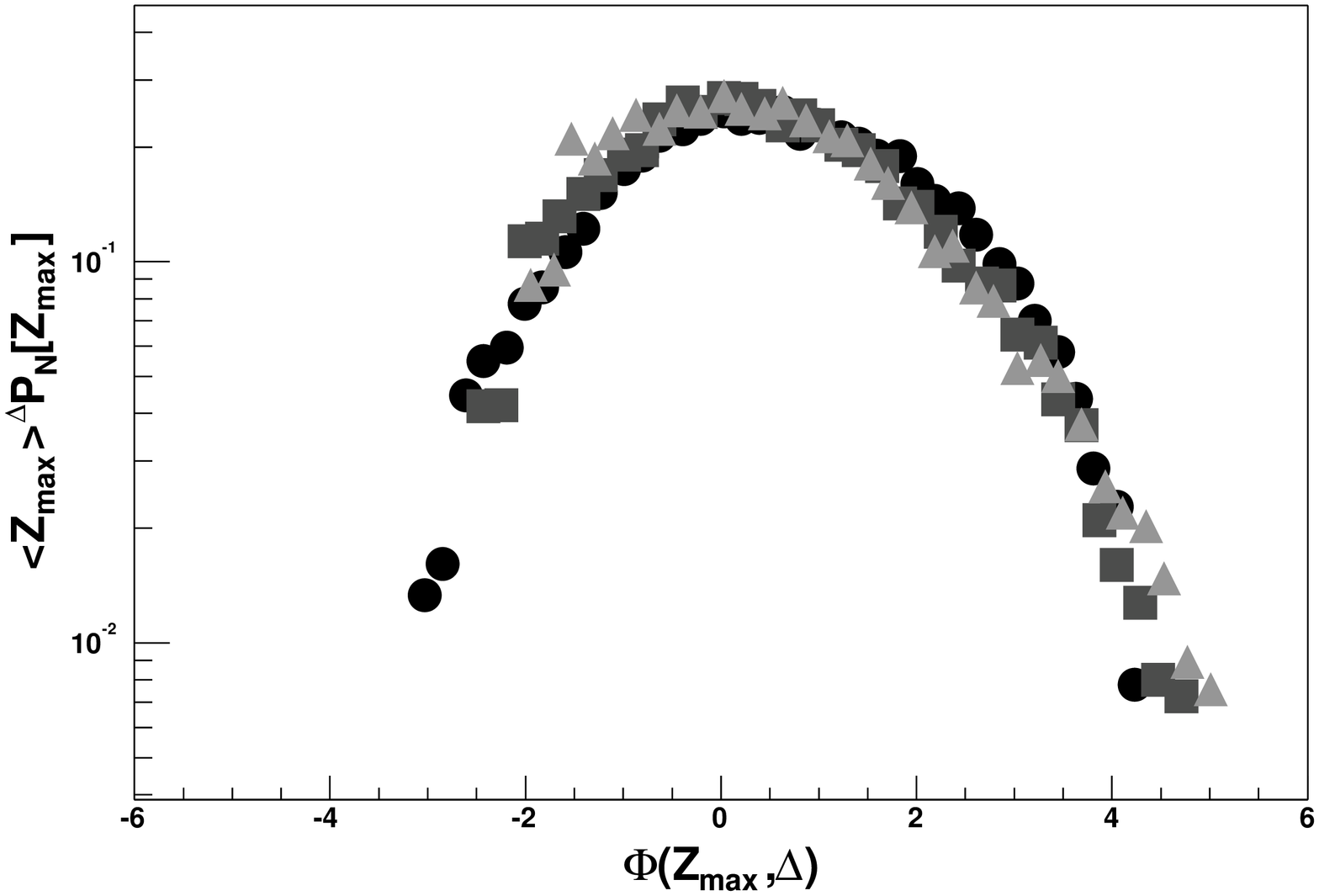}
\end{minipage}
\vspace{-12mm}
\caption{Left panel: $Z_{\rm max}$ distributions for $^{124}$Sn 
projectile spectator, scaled using $\Delta$=1. Right panel: 
$\Delta$ distributions for the 
same system scaled using $\Delta$=0.5. The various symbols refer to different cuts in the
 hodoscope multiplicity.}
\label{dsca}
\end{figure}
An additional selection has, moreover, been applied to the data by defining
two cuts on the asymmetry variable: 
$\eta_{\rm asy}>0.5$ and $\eta_{\rm asy}<0.3$ corresponding to the {\em liquid} and {\em gas} phases
respectively. 
The correlation between the mean value and the fluctuation of $Z_{\rm max}$ 
is shown in fig.~\ref{dsca1}. 
The dotted lines represent results of the fits providing $\Delta \approx$ 1 for high 
excitation energy (left-hand side of fig.~\ref{dsca1}) and $\Delta \approx$ 0.5
for low excitation energy (right-hand side of fig.~\ref{dsca1}).
The form of the $Z_{\rm max}$ distributions also evolves with excitation energy: 
it is nearly Gaussian in the $\Delta$=1/2 regime (right panel of
fig.~\ref{dsca}) beside for $\Delta$=1 the shape is asymmetric with a near-exponential tail for large positive values 
of the scaling variable, $z_{(\Delta)}$ (left panel of fig.~\ref{dsca}): 
this distribution approaches that of the modified
Gumbel-distribution~\cite{frank}. \\
Interestingly enough the transition from the {\it ordered} to the {\it disordered} regimes
is observed for the same impact parameter bin, for which the appearance of a bimodal distribution 
has been observed (see fig.~\ref{bimod}). Similar results have been obtained in the
study of the {\it quasi}-projectile decay at lower energies~\cite{iwm}. 
The interpretation of the $\Delta$-scaling result is still an open question. 
The two $\Delta$-scaling regimes seem to suggest 
that $Z_{\rm max}$ is behaving
like the order-parameter of a second-order phase transition. Moreover, other analyses 
of similar data have found
 evidence for
multifragmentation being a first-order phase transition, through signals of liquid-gas 
coexistence in the form, 
for example, of a
flattening of the caloric curve~\cite{joe}. 
This apparent contradiction could be understood as a finite-size effect~\cite{francy}.
The fact that many reactions exhibit a universal behavior, dependent only on 
the available energy, 
can again be an indication of a microcanonical equilibrium.\\
\\
{\small \it C.Sf. acknowledges the receipt of an Alexander-von-Humboldt fellowship. 
This work was supported by the European 
Community under contract No. RII3-CT-2004-506078 and HPRI-CT-1999-00001 and by the Polish Scientific 
Research Committee under contract No. 2P03B11023.}

\end{document}